\documentclass[prl,twocolumn,twoside,preprintnumbers,superscriptaddress,nofootinbib]{revtex4}
\usepackage{amsmath}
\usepackage{amssymb}
\usepackage{hyperref}
\usepackage{xspace}
\usepackage{slashed}
\hypersetup{
    colorlinks=true,       
    linkcolor=blue,          
    citecolor=blue,        
    filecolor=blue,      
    urlcolor=blue           
}
\usepackage{xcolor}
\usepackage{color}
\usepackage{graphicx}  %
\usepackage{amsmath}   %
\usepackage{graphics}

\newcommand{\MS}{$\overline{\text{MS}}$ }

\newcommand{\A}{\mathcal{A}}
\newcommand{\kk}{\mathbf{k}}
\newcommand{\pp}{\mathbf{p}}
\newcommand{\slashpi}{\protect{\slash\hspace{-0.5em}\pi}}
\newcommand{\Order}{\mathcal{O}}

\newcommand{\beq}{\begin{equation}}
\newcommand{\eeq}{\end{equation}}

\newcommand{\Fpi}{F_\pi}

\newcommand{\GeV}{\,\text{GeV}}
\newcommand{\MeV}{\,\text{MeV}}
\newcommand{\fm}{\,\text{fm}}
\newcommand{\mpi}{M_\pi}

\allowdisplaybreaks[1]

\begin{document}
\preprint{LA-UR-20-30355}

\title{
Toward complete leading-order predictions for neutrinoless double $\boldsymbol{\beta}$ decay
}

\author{Vincenzo Cirigliano}
\affiliation{Theoretical Division, Los Alamos National Laboratory, Los Alamos, NM 87545, USA}

\author{Wouter  Dekens}
\affiliation{Department of Physics, University of California at San Diego, La Jolla, CA 92093, USA} 

\author{Jordy de Vries}
\affiliation{Institute for Theoretical Physics Amsterdam and Delta Institute for Theoretical Physics, University of Amsterdam, Science Park 904, 1098 XH Amsterdam, The Netherlands}
\affiliation{Nikhef, Theory Group, Science Park 105, 1098 XG, Amsterdam, The Netherlands}
\affiliation{Amherst Center for Fundamental Interactions, Department of Physics, University of Massachusetts, Amherst, MA 01003, USA}
\affiliation{RIKEN BNL Research Center, Brookhaven National Laboratory,
Upton, New York 11973-5000, USA}

\author{Martin Hoferichter}
\affiliation{Albert Einstein Center for Fundamental Physics, Institute for Theoretical Physics, University of Bern, Sidlerstrasse 5, 3012 Bern, Switzerland}

\author{Emanuele Mereghetti}
\affiliation{Theoretical Division, Los Alamos National Laboratory, Los Alamos, NM 87545, USA}
 
\begin{abstract}
The amplitude for the neutrinoless double $\beta$ ($0\nu\beta\beta$) decay of the two-neutron system, $nn\to ppe^-e^-$, constitutes a key building block for nuclear-structure calculations of heavy nuclei employed in large-scale $0\nu\beta\beta$ searches. Assuming that the $0\nu\beta\beta$ process is mediated by a light-Majorana-neutrino exchange, a systematic analysis in chiral effective field theory shows that already at leading order a contact operator is required to ensure renormalizability. In this Letter, we develop a method to estimate the numerical value of its coefficient (in analogy to the Cottingham formula for electromagnetic contributions to hadron masses) and validate the result by reproducing the charge-independence-breaking contribution to the nucleon--nucleon scattering lengths. Our central result, while derived in dimensional regularization, is given in terms of the renormalized amplitude $\A_\nu(|\pp|,|\pp^\prime|)$, matching to which will allow one to determine the contact-term contribution in regularization schemes employed in nuclear-structure calculations.      
Our results  thus greatly  reduce a crucial uncertainty in the interpretation of searches for $0\nu\beta\beta$ decay.     
\end{abstract} 

\maketitle

\emph{Introduction}.---Neutrinoless double $\beta$ decay is by far the most sensitive laboratory probe of lepton number violation (LNV). Its observation would prove that neutrinos are Majorana fermions, constrain neutrino mass parameters, and provide 
 experimental validation for  leptogenesis scenarios~\cite{Furry:1939qr,Schechter:1981bd,Davidson:2008bu,Rodejohann:2011mu}.  
If $0\nu\beta\beta$ decay is caused by the exchange of light Majorana neutrinos, as we consider here, the amplitude is proportional to the ``effective'' neutrino mass $m_{\beta\beta}=\sum_i U_{ei}^2 m_i$, where the sum runs over light neutrino masses $m_i$ and $U_{ei}$ are elements of the neutrino-mixing matrix. $0\nu\beta\beta$ is a complicated process involving particle, nuclear, and atomic physics and the interpretation of experimental limits~\cite{KamLAND-Zen:2016pfg,Arnold:2016qyg,Albert:2017owj,Aalseth:2017btx,Adams:2019jhp,Agostini:2020xta}, and even more so 
of potential future discoveries, is hampered by substantial uncertainties in the calculation of hadronic and nuclear matrix elements~\cite{Avignone:2007fu,Menendez:2008jp,Vergados:2012xy,Simkovic:2013qiy,Vaquero:2014dna,Barea:2015kwa,Senkov:2015juo,Engel:2016xgb,Dolinski:2019nrj}.       

\begin{figure}[t]
 \centering
 \includegraphics[width=\linewidth,clip]{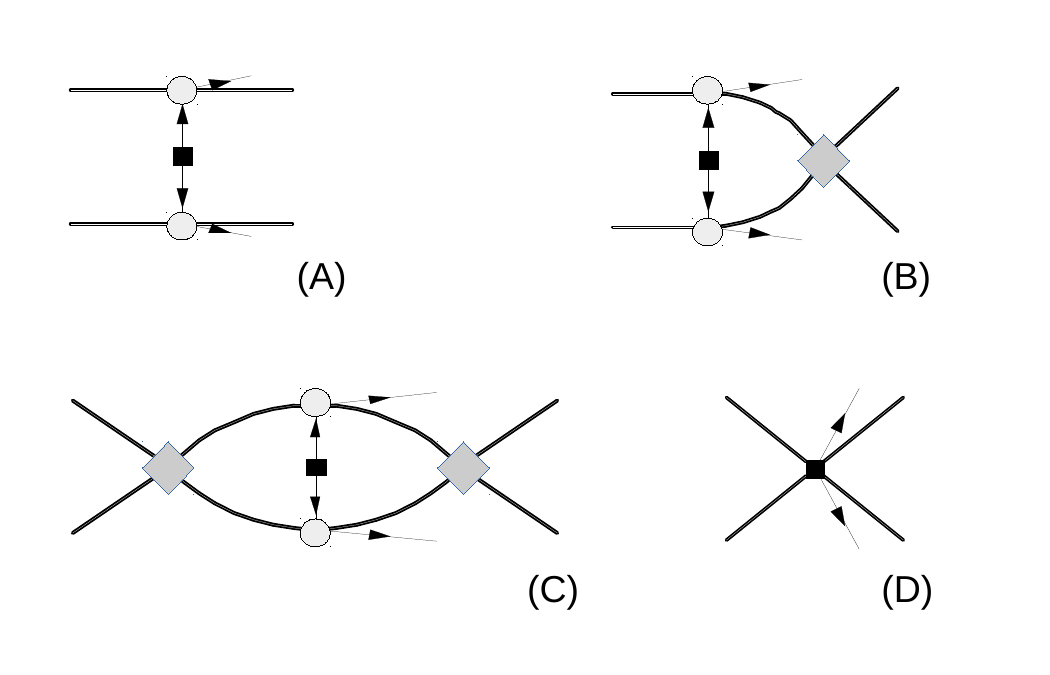}
 \caption{LO topologies for $0\nu\beta\beta$: the thick solid lines denote nucleons and the oriented ones 
 leptons (internal neutrino and  external electrons).  The squares denote LNV vertices.
 The diagrams for the electromagnetic current are obtained by replacing 
the internal neutrinos by photons, 
omitting the external electrons, and adding 
an additional topology with the internal neutrinos replaced by pions. 
In the full theory, the EFT vertices, here denoted by gray circles and diamonds, are supplemented by the appropriate form factors and scattering amplitudes that capture the momentum dependence of the elastic $N\!N$ intermediate-state contributions.
Iterations of the $N\!N$ strong Yukawa and short-range interactions (diamonds) 
are not shown as they are irrelevant for the matching analysis. 
 }
 \label{fig:diagrams}
\end{figure}

Chiral effective field theory (EFT)
\cite{Weinberg:1978kz,Weinberg:1990rz,Weinberg:1991um,Kaplan:1996xu,Kaplan:1998tg,Kaplan:1998we}  
plays a key role in addressing these uncertainties.  Nuclear structure, \textit{ab-initio} calculations based on chiral-EFT interactions~\cite{Epelbaum:2008ga,Machleidt:2011zz,Hammer:2019poc} 
have recently become available for some  phenomenologically relevant nuclei~\cite{Yao:2019rck,Belley:2020ejd,Novario:2020dmr} and the issue of $g_A$ quenching in single $\beta$ decays has been resolved as a combination of two-nucleon 
weak currents 
and strong correlations in the nucleus~\cite{Towner:1987zz,Pastore:2017uwc,Gysbers:2019uyb}. In addition, the few-nucleon amplitudes used as input in nuclear structure calculations have been scrutinized in chiral EFT
for various sources of LNV~\cite{Prezeau:2003xn,Menendez:2011qq,Cirigliano:2017djv,Cirigliano:2017tvr,Pastore:2017ofx,Cirigliano:2018hja,Wang:2018htk,Cirigliano:2018yza,Cirigliano:2019vdj,Dekens:2020ttz}. 
In the context of light-Majorana-neutrino exchange, using 
naive dimensional counting, the leading contribution in the chiral-EFT expansion arises from a 
neutrino-exchange diagram, in which the LNV arises from insertion of the $\Delta L=2$ effective neutrino mass $m_{\beta \beta}$ (see diagram (A) in Fig.~\ref{fig:diagrams}). 
In analogy to the nucleon--nucleon ($N\!N$) potential itself~\cite{Kaplan:1996xu,Kaplan:1998tg,Kaplan:1998we}
and  external currents~\cite{Valderrama:2014vra}, this conclusion no longer holds when demanding manifest renormalizability of the amplitude, 
which requires the promotion of an $nn\to ppe^-e^-$ contact operator to leading order (LO)~\cite{Cirigliano:2018hja,Cirigliano:2019vdj} (see diagram (D) in Fig.~\ref{fig:diagrams}), 
encoding the exchange of neutrinos with energy/momentum  greater than the nuclear scale. 
The size of this contact operator is currently unknown, leading to an additional source of uncertainty in the interpretation of $0\nu\beta\beta$ decays besides the nuclear-structure ones. 
In this Letter  we present a first estimate of the complete $nn\to pp e^-e^-$ amplitude including this contact-term contribution. 
For related progress toward a calculation of this amplitude  based on lattice gauge theory, 
we refer to the recent literature~\cite{Feng:2018pdq,Tuo:2019bue,Cirigliano:2020yhp,Detmold:2020jqv,Feng:2020nqj,Davoudi:2020ngi,Davoudi:2020gxs} (see Ref.~\cite{Richardson:2021xiu} for a large-$N_c$ analysis).

\begin{figure}[t]
 \centering
 \includegraphics[width=\linewidth,clip]{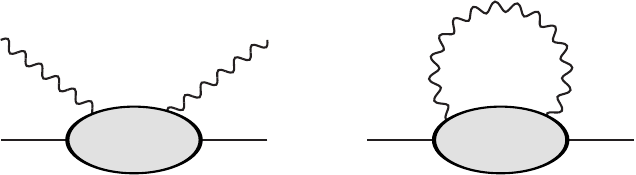}
 \caption{Forward scattering amplitude (left) and self-energy contraction (right). 
 The solid line refers to the hadronic states (pion, nucleon, two-nucleon), 
 the gray blob to the nonperturbative amplitude, 
 and the wiggly lines to the massless 
mediator  attached to the currents (photon or neutrino).}
 \label{fig:self_energy}
\end{figure}

The 
hadronic part of the 
light-Majorana-neutrino-exchange amplitude has the structure 
\beq
{\cal A}_\nu  \propto  
\int \frac{d^4k}{(2\pi)^4} \frac{g_{\mu \nu}}{k^2 + i \epsilon}  
\int d^4x\, e^{i k \cdot x} 
 \langle pp | 
T\{j_\text{w}^\mu(x) j_\text{w}^\nu(0)\}
|nn \rangle 
\label{eq:M1}
\eeq
and is ultimately determined by the 
two-nucleon 
matrix element of the time-ordered product 
$T\{j_\text{w}^\mu(x) j_\text{w}^\nu(0)\}$
of two weak currents.  
Similar matrix elements arise in electromagnetic contributions to hadron masses or scattering processes, replacing the weak current by the electromagnetic current $j_\text{em}^\mu(x)$.
In the case of hadron masses, the Cottingham formula~\cite{Cottingham:1963zz,Harari:1966mu}   relates the electromagnetic splitting to the contraction of the forward Compton scattering amplitude  with a  massless propagator, see Fig.~\ref{fig:self_energy} and Eq.~\eqref{eq:M1}. In this case, at least the elastic contribution to the mass (for which the hadronic intermediate state is the same as the external ones) follows unambiguously from the contraction of the scattering amplitude.
Since this is precisely the same structure as required for the light-Majorana-neutrino-exchange contribution to the $0\nu\beta\beta$ decay $nn\to pp e^-e^-$, the novel idea of this Letter is to constrain the corresponding amplitude by generalizing the Cottingham approach to the two-nucleon system, and then determine the contact-term contribution by matching to chiral EFT. 

In the application of the Cottingham approach to the pion and nucleon mass difference, the by far dominant contribution arises from elastic intermediate states: the pion-pole contribution gives more than $80\%$ of the pion mass difference~\cite{Ecker:1988te,Bardeen:1988zw,Donoghue:1993hj,Baur:1995ig,Donoghue:1996zn}, and the nucleon pole provides the bulk of the electromagnetic part of the proton--neutron mass difference $m_{p-n}^\text{el}=0.75(2)\MeV$. In this case, there is a tension between the estimate of the inelastic contributions in lattice QCD, $m_{p-n}^\text{inel}=0.28(11)\MeV$~\cite{Borsanyi:2014jba,Brantley:2016our,Horsley:2019wha}, and from nucleon structure functions, $m_{p-n}^\text{inel}=-0.17(16)\MeV$~\cite{Gasser:2020mzy,Gasser:2020hzn,Gasser:2015dwa,Gasser:1974wd}, but in either case the elastic estimate is accurate at the $30\%$ level. 

The main complication in the generalization to $0\nu\beta\beta$ decay is due to the two-particle nature of initial and final states and the ensuing proliferation of kinematic variables and scalar functions in a Lorentz decomposition of the amplitude. Accordingly, we do not attempt a strict derivation of the elastic contribution via a dispersion relation, but include the most important intermediate states in close analogy to the results for the pion and nucleon Cottingham formula, as described in more detail below. To assess the validity and accuracy of the approach, 
we also consider the two-nucleon matrix element  with two electromagnetic currents, 
which controls charge independence breaking (CIB) in the $N\!N$ scattering lengths. 
In this case, comparison with data 
allows us to confirm the expectation  
of an accuracy around $30\%$ if only elastic contributions are kept, as suggested by the proton--neutron mass difference. 
A determination at this level already has a major impact in bounding the size of the 
contact-term contribution to $0\nu\beta\beta$ decay.

\emph{Matching procedure}.---The integration over the neutrino momentum $k$ in Eq.~\eqref{eq:M1}  can be split into several regions. Given the nonrelativistic nature of the process, we always perform the $k^0$ integral via the residue theorem, and analyze the relevant momentum regions in terms of the space-like modulus $|\kk|$. We introduce a low-energy region $|\kk|<\Lambda_{\chi}$, a hard region $|\kk|>\Lambda$, and an intermediate region $\Lambda_\chi < |\kk| < \Lambda$, where $\Lambda$ denotes the scale at which an operator product expansion (OPE) becomes applicable and $\Lambda_\chi$ the breakdown scale of chiral EFT. The basic idea in generalizing the Cottingham approach to the $N\!N$ system then amounts to interpolating between the low-energy EFT and the high-energy OPE constraints, by using information on the momentum dependence of the pion and nucleon form factors as well as the $N\!N$ scattering amplitude. This strategy captures the analog of the dominant elastic contributions to the Cottingham formula for the pion and nucleon mass differences, but does introduce some model dependence in the intermediate region. To estimate the uncertainty it is thus critical to be able to validate the approach with data, as we will demonstrate below in terms of CIB in the $N\!N$ scattering lengths. In practice, we express our result for the full amplitude  as \looseness-1
\begin{align}
\A_\nu^\text{full}  &= 
 \int_0^\infty \,  d |\kk|  \ a^\text{full}  (|\kk|) = 
 \A^< + \A^>,\\
\A^<  &=    \int_0^{\Lambda} \,  d |\kk|  \ a_< (|\kk|),\qquad
\A^>  =  \int_{\Lambda}^\infty \,  d |\kk|  \ a_> (|\kk|),\notag 
\end{align}
where $\A^<$ subsumes the low- and intermediate-momentum regions, $\A^>$ denotes the short-distance contribution, and the two are separated by the scale $\Lambda$ where the OPE behavior is expected to set in. In the final step, we match the EFT description to the full amplitude
\beq
\A_\nu^\text{EFT}=\A^< + \A^>,
\eeq
which determines the contact-term contribution in the \MS scheme employed in the EFT calculation.

\emph{Pion mass difference}.---To illustrate this matching procedure, we first reformulate the Cottingham formula for the pion mass difference in terms of our matching variable $|\kk|$ instead of the Wick-rotated four-momentum $k_E$ typically considered in the literature~\cite{Ecker:1988te,Bardeen:1988zw,Donoghue:1993hj,Baur:1995ig,Donoghue:1996zn}. 
In this case, the matching proceeds in terms of the low-energy constant $Z$, which determines, at LO, the pion mass difference
\beq
\label{Z_pheno}
Z=\frac{M_{\pi^\pm}^2-M_{\pi^0}^2}{2e^2 F_\pi^2}=0.81.
\eeq
The elastic contribution to the Cottingham formula gives
\begin{align}
\label{Z_Cottingham}
Z^\text{el} &= \frac{3 i}{2 \Fpi^2} \, \int \frac{d^4k}{(2 \pi)^4} \ \frac{[F_\pi^V(k^2)]^2}{k^2 + i \epsilon}\notag\\
&=\frac{3}{32\pi^2\Fpi^2}\int_0^\infty d k_E^2 \big[F_\pi^V(-k_E^2)\big]^2=\frac{3M_V^2}{32\pi^2\Fpi^2}, 
\end{align}
where, for the pion vector form factor, the simple approximation 
$F_\pi^V(k^2)=M_V^2/(M_V^2-k^2)$ with $M_V=M_\rho$ has been inserted. In our matching procedure, the low-energy contribution is instead identified by introducing a cutoff in $|\kk|$, which leads to
\beq
Z^<=\frac{3}{16 \pi^2 \Fpi^2}
\int_0^\Lambda d |\kk| \,  |\kk| \frac{(\omega_V-|\kk|)^2(2\omega_V+|\kk|)}{\omega_V^3},
\eeq
with $\omega_V=\sqrt{M_V^2+|\kk|^2}$. For $\Lambda\to \infty$ this expression agrees with Eq.~\eqref{Z_Cottingham}. For our application, the $\rho$-pole approximation for $F_\pi^V$ is sufficient, but could be extended by introducing a dispersive representation~\cite{Colangelo:2018mtw}, whose Cauchy kernel would be treated in analogy to the vector-meson propagator above, via the residues in the $k^0$ integration. Second, we find
\beq
Z^> =  \frac{3 \alpha_s (\mu) g_{LR}^{\pi \pi} (\mu)}{16 \pi}  \int_{\Lambda}^{\infty}   d |\kk|  \, \frac{1}{|\kk|^3}, 
 \eeq
with coefficient $g_{LR}^{\pi \pi} =(4\pi\Fpi)^2\bar g_{LR}^{\pi \pi}$, $\bar g_{LR}^{\pi \pi}=8.2$ at \MS scale $\mu=2\GeV$~\cite{Nicholson:2018mwc} (see Refs.~\cite{Collins:1978hi,Hill:2016bjv} for the OPE contribution in the nucleon case). At scale $\Lambda=2\GeV$ we find for the sum $Z=Z^< + Z^> = 0.60 + 0.03 = 0.63$, which for $\Lambda\to\infty$ approaches $Z=0.67$. The deficit to Eq.~\eqref{Z_pheno} is understood in terms of inelastic contributions from axial-vector intermediate states~\cite{Ecker:1988te,Bardeen:1988zw,Donoghue:1993hj,Baur:1995ig,Donoghue:1996zn}, which provides another estimate of the error incurred by only considering elastic contributions.  \looseness-1

\begin{figure}[t]
 \centering
 \includegraphics[width=\linewidth,clip]{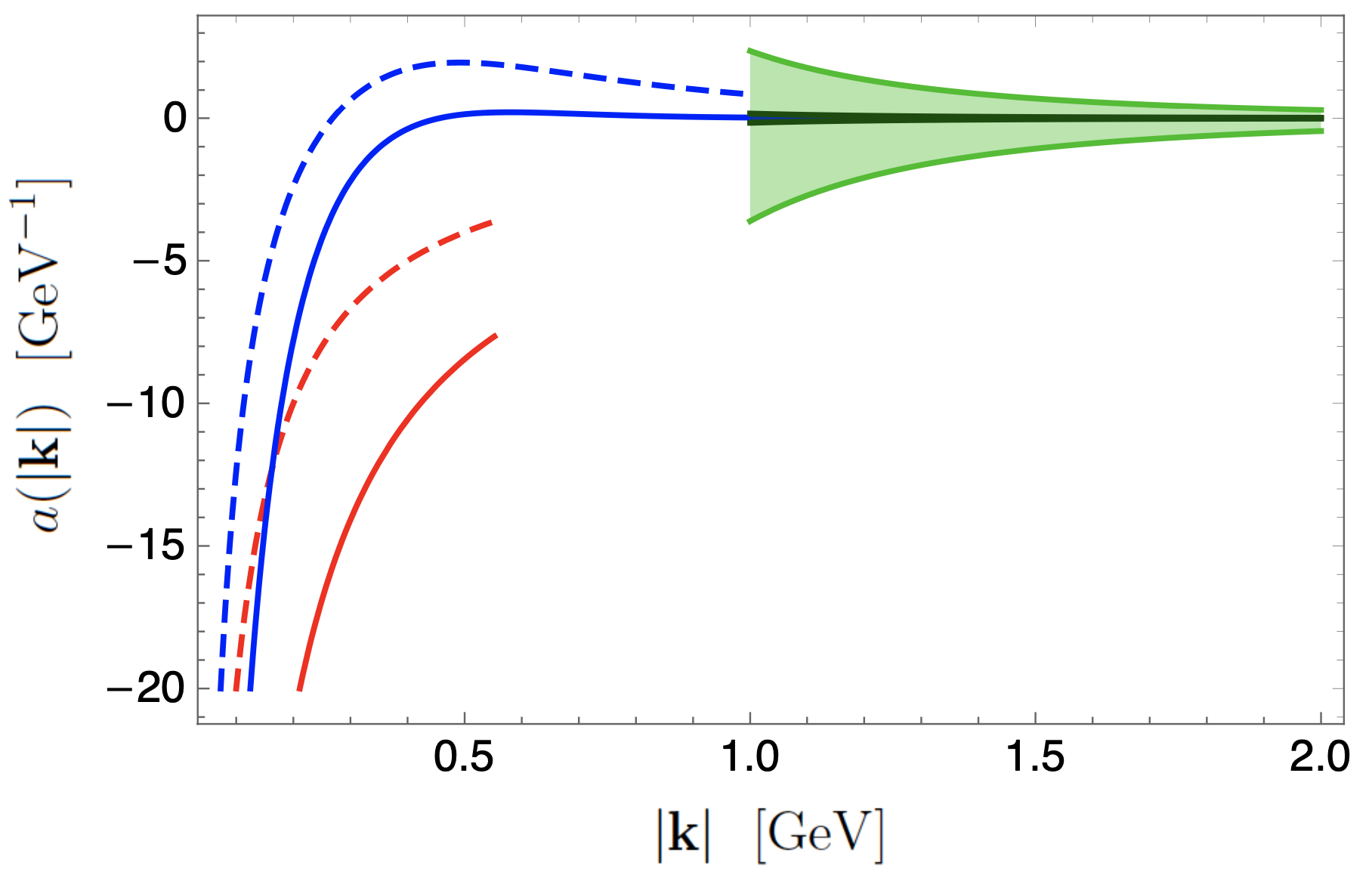}
 \caption{Integrand functions 
 defined in Eq.~\eqref{matching_integrands},
appearing in the matching relations \eqref{matching} and \eqref{matching_integrals} for  
the LNV coupling   $\tilde{\cal C}_1$: 
 $a_\chi (|\kk|)$  (solid red line extending to $|\kk| = 4 M_\pi$), 
$a_< (|\kk|)$   (solid blue line), 
and  $a_> (|\kk|)$  (dark-green thin band obtained by taking the range  $\bar g_1^{NN}  \in [-10,10]$). 
The dashed red/blue lines and light green band correspond to 
the integrands $a_{\chi,< ,>}(|\kk|)$  entering the matching relation for the CIB coupling  $\tilde{\cal C}_1+ \tilde{\cal C}_2$. 
The light green band corresponds to the range  $\bar{g}_{LR}^{NN} \in [-10,10]$. 
}
 \label{fig:integrand}
\end{figure}

\emph{Contact term in $0\nu\beta\beta$ decay}.---The $nn \to ppe^-e^-$  amplitude in chiral EFT takes the form
\beq
\A_\nu^\text{EFT}=\A_A+\A_B+\A_C+\A_D,
\eeq
where the four terms correspond to the topologies in Fig.~\ref{fig:diagrams}, and renormalization of the divergence in $\A_C$ requires the LO contact term $\A_D$. 
For the matching, only these latter two topologies become relevant. 
In particular,  only the ultraviolet singular part of the $C$  topology---i.e., the one involving noninteracting two-nucleon propagators---enters the matching condition, which can be expressed in terms of 
dimensionless amplitudes as
\beq
\label{matching}
\bar \A_C^{<, \text{sing}}+ \bar \A_C^{>}=
\bar \A_C^{\text{sing}}(\mu_\chi)+2\tilde {\cal C}_1(\mu_\chi).
\eeq
The left-hand side refers to the full amplitude, separated into momentum  regions in analogy to $Z^{\lessgtr}$ above, 
while the right-hand side gives the amplitude in chiral EFT including the contact term $\tilde {\cal C}_1$ at \MS scale $\mu_\chi$ (we use here the notation of Ref.~\cite{Cirigliano:2019vdj}). 
The explicit expressions are
\begin{align}
\label{matching_integrals}
\bar\A^{<, \text{sing}}_{C}  &=  \int_0^{\Lambda} \  d |\kk| \ a_< ( | \kk |),\quad\!  
\bar\A^{>}_{C} =   \int_{\Lambda}^\infty   \  d |\kk| \ a_> ( | \kk |),\notag \\
 \bar\A^\text{sing}_{C}  (\mu_\chi) & =   - \frac{1+2 g_A^2}{2} + 
 \int_{0}^{\mu_\chi}   \  d |\kk| \ a_\chi ( | \kk |),
\end{align}
with integrands
\begin{align}
\label{matching_integrands}
a_< ( | \kk |)  & =  - \frac{r(|\kk|)}{|\kk|} \,  \theta(|\kk| - 2 |\pp|)\notag\\
&\times\bigg[
\big[g_V  (\kk^2)\big]^2 + 2 \big[g_A( \kk^2 )\big]^2 +   \frac{\kk^2 \big[g_M (\kk^2)\big]^2}{2 m_N^2}
\bigg]  
,\notag
\\
a_> ( | \kk |)    &=
 \frac{3 \alpha_s(\mu)}{\pi} \ \bar{g}_1^{NN}(\mu) \ \frac{F_\pi^2}{ |\kk|^3},\notag\\
  a_\chi ( | \kk |)   &=  - (1+2 g_A^2)    \ 
\frac{1}{ |\kk|} \,  \theta(|\kk| - 2 |\pp|),
\end{align}
where $g_{V,A,M}(\kk^2)$ refers to the appropriate nucleon form factors in analogy to $F_\pi^V(k^2)$ above, $\bar{g}_1^{NN}(\mu)$  
is the two-nucleon matrix element of the local operator controlling   
the short-distance behavior of  $T\{j_\text{w}^\mu(x) j_\text{w}^\nu(0)\}$,  
 and $\pp$ denotes the momentum of the incoming $nn$ pair. In addition, compared to the pion mass example, 
there is a new source of momentum dependence originating from the $N\!N$ scattering amplitude itself, parameterized here in terms of $r(|\kk|)$. At LO in chiral EFT $r^\text{LO}(|\kk|)=1$, with corrections that, in pionless EFT, can be identified with the effective range $r_0$, $r^\text{NLO}_\slashpi ( |\kk|) = 1 - r_0 |\kk|/\pi$. In practice, we have evaluated $r(|\kk|)$ using next-to-leading-order chiral EFT as well as the $N\!N$ potentials from Refs.~\cite{Reid:1968sq,Wiringa:1994wb,Kaplan:1999qa}, see Ref.~\cite{Cirigliano:2021qko} for more details. 
For the nucleon form factors simple dipole parameterizations are sufficient, with the main uncertainty arising from the axial-vector scale, which we take as $\Lambda_A=1.0(2)\GeV$ to match the uncertainty for the axial radius quoted in Ref.~\cite{Hill:2017wgb}.      
The 
matrix element   
$\bar{g}_1^{NN}(\mu)$, expected to be $\Order(1)$, is presently unknown, but  in view of the large 
corresponding pion matrix element $\bar g_{LR}^{\pi \pi} = 8.2$  we take   $\bar g_1^{NN} \in[-10,10]$. 
The  impact  on the numerical analysis remains minor, reflecting the stability of the result upon variation of $\Lambda$, as long as $\Lambda > 1\GeV$. 
The functions $a_{\chi,<,>} (|\kk|)$  determining the $N\!N$ (elastic) contribution to the amplitude are shown in Fig.~\ref{fig:integrand}.   
Finally, the main uncertainty is expected to arise from inelastic contributions. To estimate their impact, we have considered the simplest diagram with an $N\!N\pi$ cut, which affects the contact term at the level of $0.1$--$0.35$, motivating an inelastic uncertainty of $0.5$. Taking everything together, we quote
\beq
\label{C1}
 \tilde {\cal C}_1 (\mu_\chi = \mpi)  
= 1.32(50)_\text{inel}(20)_{r}(5)_\text{par}
 =  1.3(6)
\eeq
as our main result for the contact term at \MS scale $\mu_\chi=\mpi$. The uncertainties refer to the inelastic contributions, $r(|\kk|)$, and parametric uncertainties (nucleon form factors and $\bar{g}_1^{NN}$), respectively. The final uncertainty is dominated by 
inelastic effects  and implies a relative precision of  (20--30)\%  on  the renormalized  singular  amplitude  $\bar \A_C^\text{sing} + 2\tilde{\cal C}_1$ 
in Eq.~\eqref{matching} at $|\pp|\sim(20\text{--}30)\MeV$---in line with the expectation from the Cottingham analyses of pion and nucleon masses discussed above. 
Note that  this  translates into a smaller relative error on the total amplitude $\A_\nu^\text{EFT}$.

\emph{Charge independence breaking}.---The LNV contact term $\tilde {\cal C}_1$ corresponds to the  
insertion of two left-handed weak currents in Eq.~\eqref{eq:M1}. 
The insertion of two (vector) electromagnetic currents 
generates  in chiral EFT a  new contact term---denoted by $\tilde {\cal C}_1+\tilde {\cal C}_2$---that contributes to  CIB in $N\!N$ scattering processes~\cite{Cirigliano:2018hja,Cirigliano:2019vdj}.  
The calculation of  $\tilde {\cal C}_1+\tilde {\cal C}_2$ 
proceeds along similar lines to the one of $\tilde{\cal C}_1$, 
but is further complicated by the pion-exchange contribution, in which the external photon currents couple to the virtual pion. We treat this intermediate-state Compton scattering amplitude in analogy to the discussion of the pion Cottingham formula above, which, for on-shell pions, amounts to isolating the pion pole in a dispersion relation~\cite{Colangelo:2015ama,Colangelo:2017qdm,Colangelo:2017fiz}, and thus corresponds to our strategy of evaluating the elastic contributions. 
The resulting matching relation becomes analogous to Eqs.~\eqref{matching}--\eqref{matching_integrands}, but includes, in addition to the appropriately amended $N\!N$ pieces, a $\pi\pi$ contribution from the pion-exchange diagram.
The matching is illustrated in Fig.~\ref{fig:integrand} and we refer to Ref.~\cite{Cirigliano:2021qko} for the explicit expressions. 
Numerically, we obtain
\begin{align}
\label{C1C2}
(\tilde{\cal C}_1 + \tilde{\cal C}_2) (\mu_\chi =\mpi)
&=2.9(1.1)_\text{inel}(0.3)_{r}(0.3)_\text{par}\notag\\
&= 2.9(1.2).
\end{align}
The assigned inelastic error corresponds to a relative error of about 50\% in the singular $N\!N$ electromagnetic amplitude 
at $|\pp| \sim 25\MeV$,  larger than the 30\% in the weak amplitude.  
In addition to the new class of pion-exchange diagrams, this is motivated by the observation that now
the parametric error becomes more sizable, almost exclusively due to the $N\!N$ short-distance coupling $\bar g_{LR}^{NN}$ varied within $[-10,10]$ and the scale $\Lambda$ between $2$ and $4\GeV$. Indeed, if $\Lambda$ were decreased to values as low as $1\GeV$ and thus into the energy region where the applicability of the OPE becomes questionable and inelastic effects important, a variation around $1.0$ would be obtained.
Numerically, the $\pi\pi$ contribution dominates, yielding $(\tilde{\cal C}_1 + \tilde{\cal C}_2) (\mu_\chi =\mpi)|_{\pi\pi}=2.4$, while the effect from the most uncertain region, $(\tilde{\cal C}_1 + \tilde{\cal C}_2) (\mu_\chi =\mpi)|_{|\kk|\in[0.4,1.5]\GeV}=0.55$, falls safely within our uncertainty estimate.
\looseness-1

The result~\eqref{C1C2} already compares quite well to the phenomenological determination $(\tilde{\cal C}_1 + \tilde{\cal C}_2) (\mu_\chi =\mpi)=5.0$ from Ref.~\cite{Cirigliano:2019vdj}. However, since the contact term is scale and scheme dependent,
 it is more appropriate to compare directly observables calculated based on Eq.~\eqref{C1C2}. To this end, we first note that within LO chiral EFT the scattering lengths $a_{nn}$, $a_{np}$, and $a_{pp}^C$ (the latter defined in the modified effective range expansion to account for Coulomb effects~\cite{Bethe:1949yr,Jackson:1950zz,Kong:1999sf}) can be mapped onto contact terms for each channel
\begin{align}
\tilde C_{np} &=  \tilde C + \frac{e^2}{3} \left( \tilde {\cal C}_1 + \tilde {\cal C}_2 \right),\notag \\
\tilde C_{nn/pp} &= \tilde C - \frac{e^2}{6} \left( \tilde{\cal C}_1 + \tilde {\cal C}_2 \right)  \pm \frac{1}{2}  \tilde {\cal C}_\text{CSB},
\end{align}
where $\tilde C$ denotes the isospin-symmetric combination, $\tilde {\cal C}_1 + \tilde {\cal C}_2$ the CIB contribution, and $\tilde {\cal C}_\text{CSB}$ a charge-symmetry-breaking term~\cite{Epelbaum:1999zn}. To test our prediction for $\tilde {\cal C}_1 + \tilde {\cal C}_2$, we can thus use two observables to determine $\tilde C$ and $\tilde {\cal C}_\text{CSB}$, and then predict the third based on Eq.~\eqref{C1C2}. We choose
\beq
\label{aCIB}
a_\text{CIB}=\frac{a_{nn}+a_{pp}^C}{2}-a_{np}=10.4(2)\fm,
\eeq
which would isolate the CIB contribution if $N\!N$ scattering were perturbative and Coulomb interactions absent, and we have used the empirical values $a_{pp}^C=-7.817(4)\fm$~\cite{Bergervoet:1988zz,Reinert:2017usi}, $a_{np}=-23.74(2)\fm$~\cite{Klarsfeld:1984es,Machleidt:2000ge}, $a_{nn}=-18.9(4)\fm$~\cite{Chen:2008zzj}. From Eq.~\eqref{C1C2} we find $a_\text{CIB}=15.5^{+4.5}_{-4.0}\fm$, in good agreement with Eq.~\eqref{aCIB}, given that additional uncertainties from higher chiral orders could be attached. We thus conclude that the comparison to CIB validates our approach at the level of (30--50)\%, and that our uncertainty estimates are realistic.

\emph{Outlook}.---Armed with our determination of  $\tilde {\cal C}_1 $  in the \MS scheme, it becomes possible to unambiguously determine the $nn\rightarrow ppe^-e^-$ amplitude $\A_\nu$ at low energies to LO in chiral EFT. 
While dimensional regularization with minimal subtraction 
is a convenient scheme for our matching strategy, it is rarely used in nuclear calculations. Since amplitudes are observables and thus scheme independent, the LNV contact term $\tilde {\cal C}_1 $
can be obtained in any other scheme, for instance in momentum- or coordinate-space cutoff schemes often applied in the \textit{ab-initio} few-body community \cite{Epelbaum:2008ga,Machleidt:2011zz,Hammer:2012id,Hammer:2019poc,Hebeler:2020ocj}, by fitting to our synthetic 
data\footnote{The amplitude $\mathcal{A}_{\nu}$ is related to the $S$-matrix element 
for the process 
$n( \pp) \ n (- \pp) \to p ( \pp^\prime) \ p (-  \pp^\prime)  \ e (   \pp_{e}) \ e ( -\pp_{e} ) $
by  
$S_{\nu} = i  (2 \pi)^4 \, \delta^{(4)}(p_f - p_i) \,  ( 4  G_F^2 V_{ud}^2  m_{\beta \beta} \   \bar{u}_L   (\pp_e)  u_L^c (-\pp_e)  )    \, {\mathcal A}_{\nu}$. 
Various choices of $|\pp|$ and $|\pp^\prime|$ are possible, see Ref.~\cite{Cirigliano:2021qko} for more details.
} 
\begin{equation}
    \A_\nu (|\pp|, |\pp^\prime|)     e^{-i (\delta_{^1S_0}(|\pp|) + \delta_{^1S_0}( |\pp^\prime|))} = -0.0195(5)\MeV^{-2}, 
    \label{eq:sdata}
\end{equation}
where $|\pp |=25\MeV$  ($|\pp^\prime |=30\MeV$) is the neutron (proton) momentum in the center-of-mass frame.  

Recent years have seen great progress in \textit{ab-initio} calculations of  $0\nu\beta\beta$ decay rates of light nuclei~\cite{Cirigliano:2019vdj, Yao:2019rck, Belley:2020ejd, Novario:2020dmr, Yao:2020olm}, ranging from ${}^6$He to the experimentally relevant ${}^{48}$Ca and  ${}^{76}$Ge, in each case starting from microscopic chiral nuclear forces. However, these decay rates only include the long-distance neutrino-exchange contributions and omit the $\tilde {\cal C}_1$ term. Once $\tilde {\cal C}_1$ is obtained by fitting to Eq.~\eqref{eq:sdata}, this omission can now be remedied and, for the first time, complete LO calculations can be performed of nuclear $0\nu\beta\beta$ decay rates. For even heavier nuclei such as ${}^{136}$Xe, which are still beyond the reach of \textit{ab-initio} techniques, the impact of the contact term 
should 
be studied indirectly, e.g., by comparing results for nuclei accessible to both \textit{ab-initio} methods and the respective nuclear model (see Ref.~\cite{Hoferichter:2020osn} for the same strategy in the context of the axial-vector current).

For $\A_\nu$ at  the kinematic point  chosen in Eq.~\eqref{eq:sdata} and 
 in the \MS scheme at $\mu_\chi=4 \mpi$, we find that the contact-term contribution adds destructively to the neutrino exchange at the 15\%  level,   but we stress that this is a scale- and scheme-dependent statement, with similar scales in cutoff schemes indicating different, in some cases even constructive, effects~\cite{Cirigliano:2021qko}.  
Moreover, as discussed in Refs.~\cite{Cirigliano:2018hja,Cirigliano:2019vdj}, 
while a contact term of natural size affects $\Delta I=0$ transitions such as $nn \to ppe^-e^-$ at the (10--20)\% level, its  effect is amplified to the level of 
(50--70)\% in  $\Delta I=2$ nuclear transitions due to a node in the matrix element density. 
Based on our result for $\A_\nu$, 
its effect can now be addressed  in calculations of realistic $0\nu\beta\beta$ 
nuclear transitions,  greatly reducing a 
crucial uncertainty in the interpretation of future searches for $0\nu\beta\beta$ decay~\cite{LeNoblet:2020efd,Zsigmond:2020bfx,Schmidt:2019gre,Pocar:2020zqz,Tetsuno:2020ngo,Lozza:2020xig,Gando:2020cxo}.

\begin{acknowledgments}
We thank Jon Engel, Evgeny Epelbaum, Michael   Graesser, and Bira van Kolck for discussions.
The work of VC and EM is supported  by the US Department of Energy through the Los Alamos National Laboratory. Los Alamos National Laboratory is operated by Triad National Security, LLC, for the National Nuclear Security Administration of U.S.\ Department of Energy (Contract No.\ 89233218CNA000001).
WD is supported by  U.S.\ Department of Energy Office of Science, under contract  DE-SC0009919. 
MH is supported by an Eccellenza Grant (Project No.\ PCEFP2\_181117) of the Swiss National Science Foundation. JdV is supported by the 
RHIC Physics Fellow Program of the RIKEN BNL Research Center. 
\end{acknowledgments}   

\bibliographystyle{h-physrev3} 
\bibliography{bibliography}

\end{document}